\documentclass[%
 aip,
 jmp,%
 amsmath,amssymb,
 preprint,%
]{revtex4-1}

\usepackage{dcolumn}
\usepackage{graphicx,amssymb,color}
\usepackage{gensymb}
\usepackage{bm}

\begin{document}

\begin{center}
\noindent\LARGE{\textbf{Mechanical Properties and Fracture \\ Dynamics of Silicene Membranes}}
\vspace{0.6cm}

\noindent\large{\textbf{T. Botari, $^{\ast}$ $^a$, E. Perim,\textit{$^{a}$} P. A. S. Autreto\textit{$^{a}$}, A. C. T. van Duin\textit{$^{b}$}, R. Paupitz\textit{$^{c}$} and
D. S. Galvao\textit{$^{a}$}}}\vspace{0.5cm}

\textit{$^{a}$ Instituto de F\'isica ‘Gleb Wataghin’, Universidade Estadual de Campinas, 13083-970, Campinas, SP, Brazil.}

\textit{$^{b}$ Department of Mechanical and Nuclear Engineering, Penn State University,
University Park, Pennsylvania 16801, USA.}

\textit{$^{c}$ Departamento de F\'isica - Universidade Estadual
Paulista Av.24A 1515 - 13506-900 - Rio Claro -
SP - Brazil.}

$^{\ast}$ Corresponding author: tiagobotari@gmail.com
\end{center}

%Please note that \ast indicates the corresponding author(s) but no footnote text is required. 

%\noindent\textit{\small{\textbf{Received Xth XXXXXXXXXX 20XX, Accepted Xth XXXXXXXXX 20XX\newline
%First published on the web Xth XXXXXXXXXX 200X}}}

%\noindent \textbf{\small{DOI: 10.1039/b000000x}}
%\vspace{0.6cm}
%Please do not change this text.

\noindent \normalsize{As graphene became one of the most important materials today, there is a renewed interest on others similar structures. One example is silicene, the silicon analogue of graphene. It share some the remarkable graphene properties, such as the Dirac cone, but presents some distinct ones, such as a pronounced structural buckling. We have investigated, through density functional based tight-binding (DFTB), as well as reactive molecular dynamics (using ReaxFF), the mechanical properties of suspended single-layer silicene.  We calculated the elastic constants, analyzed the fracture patterns and edge reconstructions. We also addressed the stress distributions, unbuckling mechanisms and the fracture dependence on the temperature. We analysed the differences due to distinct edge morphologies, namely zigzag and armchair.}
\vspace{0.5cm}

\section{Introduction}

\begin{figure}[h]
%\vspace*{-1.1cm}
\centerline{\includegraphics[width=1.0\linewidth]{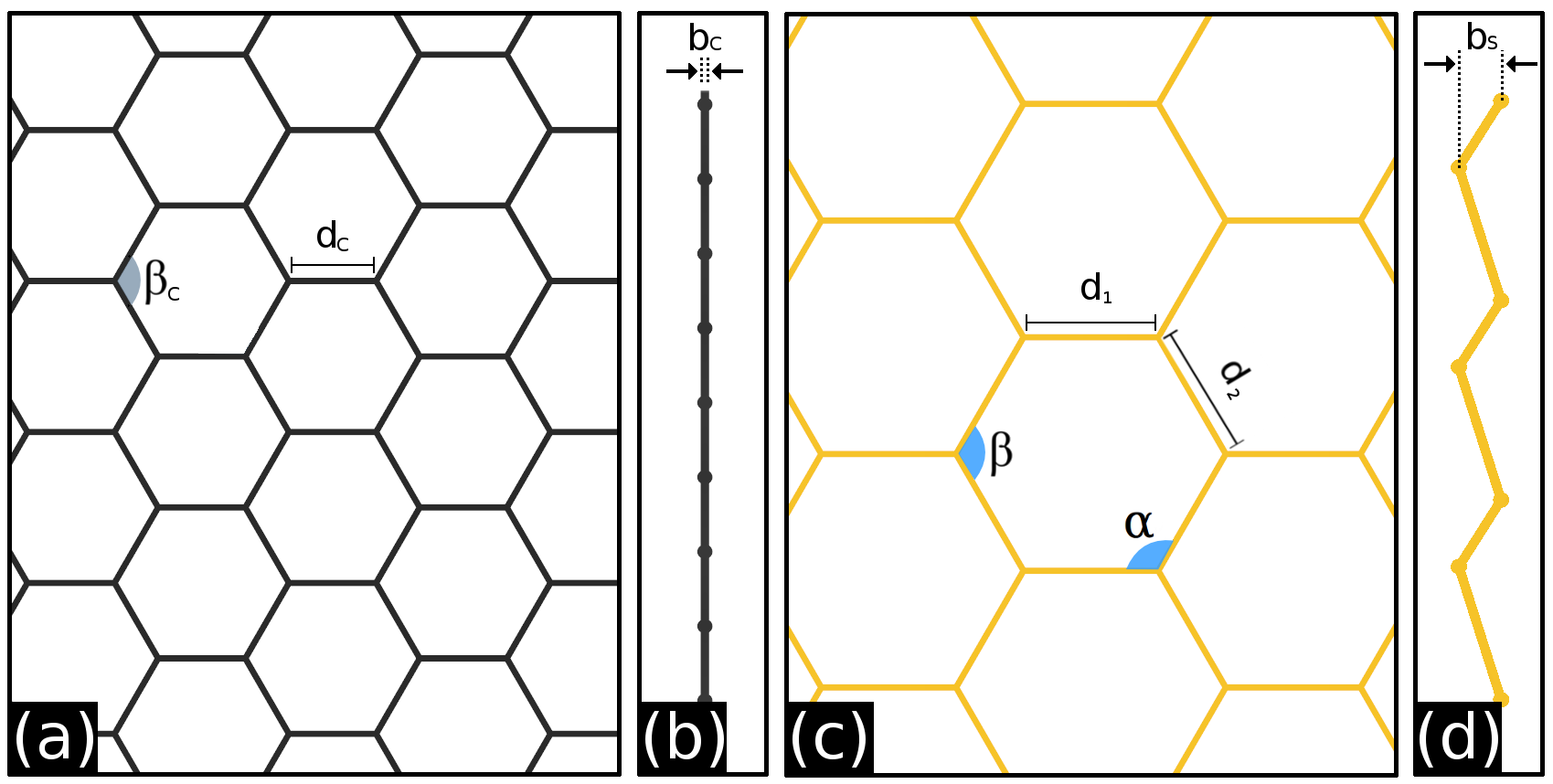}}
\caption{{(Color online) Schematic view of graphene and silicene membranes, in the same scale. (a/c) and (b/d) refer to frontal and lateral view of graphene and silicene, respectively.}}
\label{Fig1}
\end{figure}

	Carbon nanostructures have been proposed as the structural basis for a series of new technological applications. The versatility that carbon exhibits in forming different structures can be attributed to its rich chemistry, reflected on the fact that it can assume three quite distinct and different hybridization states: $sp^3$ (diamond), $sp^2$ (graphite, graphene, fullerenes and nanotubes \cite{Ref1}) an $sp$ (graphynes \cite{Ref2,Ref3,Ref4}). 
    Carbon based structures of low dimensionality exhibit extraordinary structural, thermal \cite{balandin2011thermal} and electronic \cite{knupfer2001electronic} properties. Among these structures, graphene (see Figure \ref{Fig1}) has been considered one of the most promising \cite{Ref5,Ref6,Ref7} due to its unique electronic and mechanical properties. However, its zero bandgap value hinders some transistor applications \cite{Ref7}. As a consequence, there is a renewed interest in other possible graphene-like structures, based on carbon or in other chemical elements.
	Other group IV elements, such as silicon and germanium, present a chemistry which is similar to that of carbon in some aspects, although the number of known carbon structures surpasses very much the ones based on silicon or germanium. A natural question is whether these elements could also form two dimensional honeycomb arrays of atoms, similar to graphene \cite{PRB2009}. The corresponding silicon and germanium structures were named silicene (see Figure \ref{Fig1}) and germanene \cite{Ref9}, respectively. Silicene was first predicted to exist based on \textit{ab initio} calculations in 1994 \cite{Ref8} and has been recently synthesized by different groups  \cite{Ref13,Ref14,Ref15}.  
	
	Silicene presents some properties that make it a very promising material to electronic applications. The electronic Dirac cone exhibited by graphene is also found in silicene \cite{Ref9}.     A notable difference between graphene and silicene is that while the former is completely planar, the latter presents a significant level of buckling, meaning that in silicene atoms are not in purely $sp^2$ hybridized states. This is due to the pseudo-Jahn Teller effect \cite{buckling,buckling1}, which introduces instability in high symmetry configurations, and can be exploited in some electronic applications \cite{buckling1}. It has been pointed out that puckering causes loss of the $sp^2$ character, lowering the plane stiffness and that linear atomic chains (LACs) may be formed during the fracturing process \cite{topsakalprb}. It is expected that some level of buckling should be always present in silicene, independently of the strain value \cite{PLA2012}. For hydrogenated silicenes (the so-called silicanes), it has been proposed that the buckling should decrease linearly with the strain \cite{PLA2012}.
	In the last years silicene has been object of many experimental and theoretical investigations \cite{review1, review2, review3}. Silicene nanoribbons have been experimentally produced over  Ag(110) surface \cite{review2}. Larger silicene nanosheets have been also synthesized \cite{review4}. Some of the theoretical aspects investigated include tuning of electronic properties under stress load \cite{RuiQin, topsakalprb}, transitions from semimetal to metal  \cite{EPL2012}, bandgap dependence on buckling geometries \cite{huang2013electronic}, mechanical properties \cite{ topsakalprb, QingPengl,Pei2014, botari2013mechanical}, formation of silicene between graphene layers \cite{silicene_duin2014}, the influence of defects \cite{vacancies} and chemical functionalizations \cite{silicane}. However, most studies in the literature have been based on small structures. 
	
There are several studies regarding fracture mechanisms on silicene membranes under strain\cite{roman2014mechanical, ansari2014elastic,Pei2014}. The contribution of the present work comes from an investigation of the relative importance of aspects such as edge terminations (armchair and$/$or zigzag), membrane size and temperature effects.  We have carried fully atomistic molecular dynamics (MD) simulations of silicene under dynamical strain at finite temperatures using reactive classical molecular dynamics in association with \textit{ab initio} density functional theory (DFT) and tight binding methods.

\begin{table*}[ht]
\caption{Comparison between our data and available results in the literature. $a_0$ is the lattice parameter, $\Delta$ is the buckling value, $d_{Si-Si}$ is the Silicon bond distance, $C$ is the plane stiffness, $\nu$ is the Poisson ratio and $\epsilon_c$ is the critical strain. (ZZ) and (AC) stand for Zigzag and Armchair directions, respectively. '*' means this value was estimated from the curve in Fig. 1 (g), from Topsakal and Ciraci \cite{topsakalprb}. }
\centering
%  \begin{tabular}    {p{0.15\linewidth}p{0.12\linewidth}p{0.05\linewidth}p{0.05\linewidth}p{0.07\linewidth}p{0.17\linewidth}p{0.17\linewidth}p{0.17\linewidth}}
%\resizebox{\columnwidth}{!}{
\begin{tabular}{|c|c|c|c|c|c|c|c|}
\hline
    Method & Structure & $a_0$ & $\Delta$ &  $d_{Si-Si}$ & $C$  & $\nu$ & $\epsilon_c$ \\
    Ref. & - & \AA & \AA & {\AA} & N/m & - & \\
\hline

DFT-LDA\cite{PRB2009} & Silicene & $3.83$ & $0.44$ & $2.25$ & $62$ & $0.30$ & -\\
DFT - LDA\cite{RuiQin}  & Silicene & $3.83$ & $0.42$ & $2.25$ & $63.0$ & $0.31$ & $20$\\
DFT-GGA-ours & Silicene & $3.83$ & $0.48$ & $2.28$ & - & - & - \\ 
ReaxFF-ours & Silicene & $3.80$ & $0.67$ & $2.3$ & - & - & - \\ 
SCC-DFTB-ours & Silicene & $3.87$ & $0.59$ & $2.32$ & - & - & - \\ 
DFT-GGA\cite{PSS2013} & Silicene &  & - & - & $62.4$(ZZ)/$59.1$(AC) & - & -\\
DFT-GGA\cite{topsakalprb} & Silicene & - & - & - & $62.0$ &-& -\\
DFT-GGA\cite{PLA2012} & Silicene & - & $0.45$ 
 & $2.28$   & $60.06$(ZZ)/$63.51$(AC) & $0.41$(ZZ)/$0.37$(AC)&  $14$(ZZ)/$18$(AC) \\
\hline
MD-EDIP\cite{PSS2013} & ACM/ZZM & - & - & - & $64.6/65.0$ &  & $19.5/15.5$\\
SCC-DFTB-ours & ACM/ZZM & - & $0.59$ & $2.32$ & $62.7/63.4$ &$0.30$/$0.30$ & $17$/$21$\\ 
ReaxFF-ours &ACM/ZZM& - & $0.67$ & $2.3$ &  $43.0$  & $0.28$/$0.23$ & 15/30 \\ 
DFT-GGA\cite{topsakalprb} & ACM & - & - & - & $51.0$ &-& $23$*\\
\hline
DFT-GGA\cite{PLA2012} & Silicane & - & $0.72$ & $2.36$   & $54.50$(ZZ)/$54.79$(AC) & $0.25$(ZZ)/$0.23$(AC) &  $33$(ZZ)/$23$(AC) \\
DFT-GGA\cite{silicane} & Silicane & $3.93$ & $0.72$ & $2.38$ & $52.55$  &  $0.24$ & - \\
\hline
  \end{tabular}
%  }
  \label{table1}
\end{table*}

\section{Methodology} 
\label{sec2}

We studied the structural and dynamical aspects of silicene membranes under strain and their fracture patterns using classical and quantum methods.
Equilibrium geometries were studied with three different methods, DFT, with the code Dmol3\cite{dmolcode,dmolcode1}, density functional based tight-binding method, with DFTB+ \cite{DFTBplus} and reactive classical molecular dynamics, via ReaxFF\cite{Ref18}. DFT calculations offer higher accuracy, however, in order to reliable simulate the rupturing dynamics of silicene membranes we need to use large systems, precluding the use of DFT due to the high computational costs. Thus, for the dynamical studies we used only tight-binding and reactive classical molecular dynamics calculations. The structural calculations with DFT were used in order to validate the accuracy of the other used methods.

For the DFT calculations, we used the Dmol$^3$ package as implemented on the Accelrys Materials Studio suite \cite{dmolcode,dmolcode1}. We carried out geometry optimization calculations with the Perdew-Burke-Ernzerhof (PBE) functional under the generalized gradient approximation (GGA), with all atoms free to move and full cell optimizations. The convergence criteria were $10^{-4}$ eV in energy, $0.05 eV/ $ \AA~ for the maximum force and $0.005$ \AA~ as the maximum displacement. Core electrons were explicitly treated and a double numerical plus polarization (DNP) basis set was used. Since the largest silicene membranes studied in this work contain approximately 1600 atoms, far beyond the reasonable size for a long-time all electron dynamical calculation using DFT methodology, we also used the density functional based tight-binding method (DFTB) for systems of intermediate size (hundreds of atoms) as well as a reactive force field method for systems of large size ($\sim$ 1600) atoms. 

The tight-binding calculations were carried out using the Self-Consistent Charge Density Functional based Tight-Binding (SCC-DFTB) \cite{DFTB,SCCDFTB} method, as implemented on DFTB+ \cite{DFTBplus}. The Density Functional based Tight-Binding  (DFTB) is a DFT-based approximation method and can treat systems composed by a large number of atoms. SCC-DFTB is an implementation of DFTB approach and has the advantage of using self-consistent redistribution of Mulliken charges (SCC) that corrects some deficiencies of the non-SCC standard DFTB methods \cite{SCCDFTB}. Dispersion terms are not, by default, considered in any DFTB method and were included in this work via Slater-Kirkwood Polarizable atomic model, as implemented in the DFTB+ package  \cite{DFTBplus}. 

Reactive classical molecular dynamics simulations were carried using the ReaxFF method \cite{Ref18}. ReaxFF is a reactive force field developed by van Duin, Goddard III and co-workers for use in MD
simulations of large systems. It is similar to standard non-reactive force fields, like MM3\cite{allinger1989} in which the system energy is divided into partial energy contributions associated with, 
amongst others; valence angle bending, bond stretching, and non-bonded van der Waals and Coulomb interactions. A major difference between ReaxFF and usual, non-reactive force fields, is that it 
can handle bond formation and dissociation. It was parameterized using density functional theory (DFT) calculations, being the average deviations between the heats of formation predicted by ReaxFF 
and the experiments equal to $2.8$ and $2.9$ kcal/mol, for non-conjugated and conjugated systems, respectively \cite{Ref18}. We use this force field as implemented in the Large-scale atomic/molecular massively 
parallel simulator (LAMMPS) code \cite{Ref19}. The ReaxFF force field was recently used to investigate several chemical reactions and mechanical properties of systems containing silicon atoms, 
such as the oxidation of silicon carbide \cite{Newsome2012} as well as silicene stabilized by bilayer graphene\cite{silicene_duin2014}.

Large systems consisting of semi-infinite strips under periodic boundary conditions for both 
edge morphologies, i.e., zigzag and armchair membranes (ZZM and ACM), were used to study the dynamical aspects of 
fracturing processes. Typical size of these membranes for ReaxFF simulations were 
$95$\AA ~by 
$100$ \AA, ~for armchair and zigzag edge terminated structures, respectively. Smaller 
structures were considered for DFTB$+$ calculations, in which membrane sizes were 
$28$\AA ~and $28$\AA, ~for armchair and zigzag edge terminated membranes, respectively. 
All structures were initially thermalized using molecular dynamics (MD), in a NPT ensemble with the external pressure value set to zero along the periodic direction before the stretching process is started. This procedure guaranteed the initial structures were at equilibrium dimensions and temperature, thus excluding any initial stress stemming from thermal effects. In order to simulate this stretching two different temperatures were considered, $10$K and $150$ K, controlled either by a Nose-Hoover\cite{nosehoover} or an Andersen\cite{andersen1980} thermostat as implemented on LAMMPS and DFTB+, respectively. Strain was generated by the gradual increase of the unit cell value along the periodic direction. We have used time-steps of $0.05~fs$ and a constant strain rate of $10^{-6}/fs$ was applied for the ReaxFF simulations. For the SCC-DFTB we used time-steps of $1~fs$ and applied a strain equal to $10^{-5}$ at intervals of $10 fs$, resulting in a strain rate of $10^{-6}/fs$ as in the ReaxFF case. These conditions were held fixed until the complete mechanical rupture of the membranes. Other strain rate values were tested, ranging from 10$^{-7}/fs$ to 10$^{-3}/fs$. It was verified that for a value of 10$^{-5}/fs$ or lower the results were equivalent. This strain rate is comparable to the ones used in previous studies \cite{roman2014mechanical,PSS2013,Pei2014}. Repeated runs under same conditions yielded equivalent results.

In order to obtain useful information regarding the dynamics of deformation and rupturing throughout the simulations, we calculated the virial stress  
tensor\cite{subramaniyan2008continuum,buellervonMises} which can be defined as 
\begin{equation} 
\sigma_{ij}=\frac{\sum_{k}^{N}m_{k}v_{k_i}v_{k_j}}{V}+\frac{\sum_{k}^{N}r_{k_i}\cdot f_{k_j}}{V},
\end{equation}
where N is the number of atoms, $V$ is the volume, $m$ the mass of the atom, $v$ is the velocity, $r$ is the position and $f$ the force acting on the atom.
Stress-strain curves were obtained considering the relation between the uniaxial component of stress tensor in a specific direction, namely $\sigma_{ii}$, and 
the strain defined as a dimensionless quantity which is the ratio between deformation along the considered direction and the length on the same direction\cite{buellervonMises}
\begin{equation}
\varepsilon_i = \frac{\Delta Li}{Li},
\end{equation}
where $i=1,2$ or $3$. Using this quantity it is also usefull to define the Young Modulus, 
$Y=\sigma_{ii}/\varepsilon_i$, 
and the Poisson ratio, which is the negative ratio between a transverse and an axial strain 
\begin{equation}
\nu =-\frac{d\varepsilon_i}{d\varepsilon_j},
\end{equation} 
where $i\neq j$. We also calculated a quantity which is related to the distortion state of the system, known as {\it von Mises stress}\cite{buellervonMises}, defined as
%\columnwidth

\begin{equation}
\sigma_{vm}=\sqrt{\frac{\left(\sigma_{11}-\sigma_{22}\right)^2+\left(\sigma_{22}-\sigma_{33}\right)^2+\left(\sigma_{11}-\sigma_{33}\right)^2
+6\left(\sigma_{12}^{2}+\sigma_{23}^{2}+\sigma_{31}^{2}\right)}{2}},
\label{vonMises}
\end{equation}

components $\sigma_{12}$, $\sigma_{23}$ and $\sigma_{31}$ are called shear stresses. 
von Mises stress provides very helpful information on fracturing processes because, by calculating this quantity for each timestep, it is possible to visualize the time evolution and 
localization of stress on the structure. This methodology was successfully used to investigate the mechanical failure of carbon-based nano structures such as graphene, carbon nanotubes \cite{dos2012unzipping}
and also silicon nanostructures\cite{buellervonMises}. 

\section{Results and Discussion}
\label{sec3}
\subsection{Structural investigation}

\begin{figure}[h]
%\vspace*{-1.1cm}
\centerline{\includegraphics[width=1.0\linewidth]{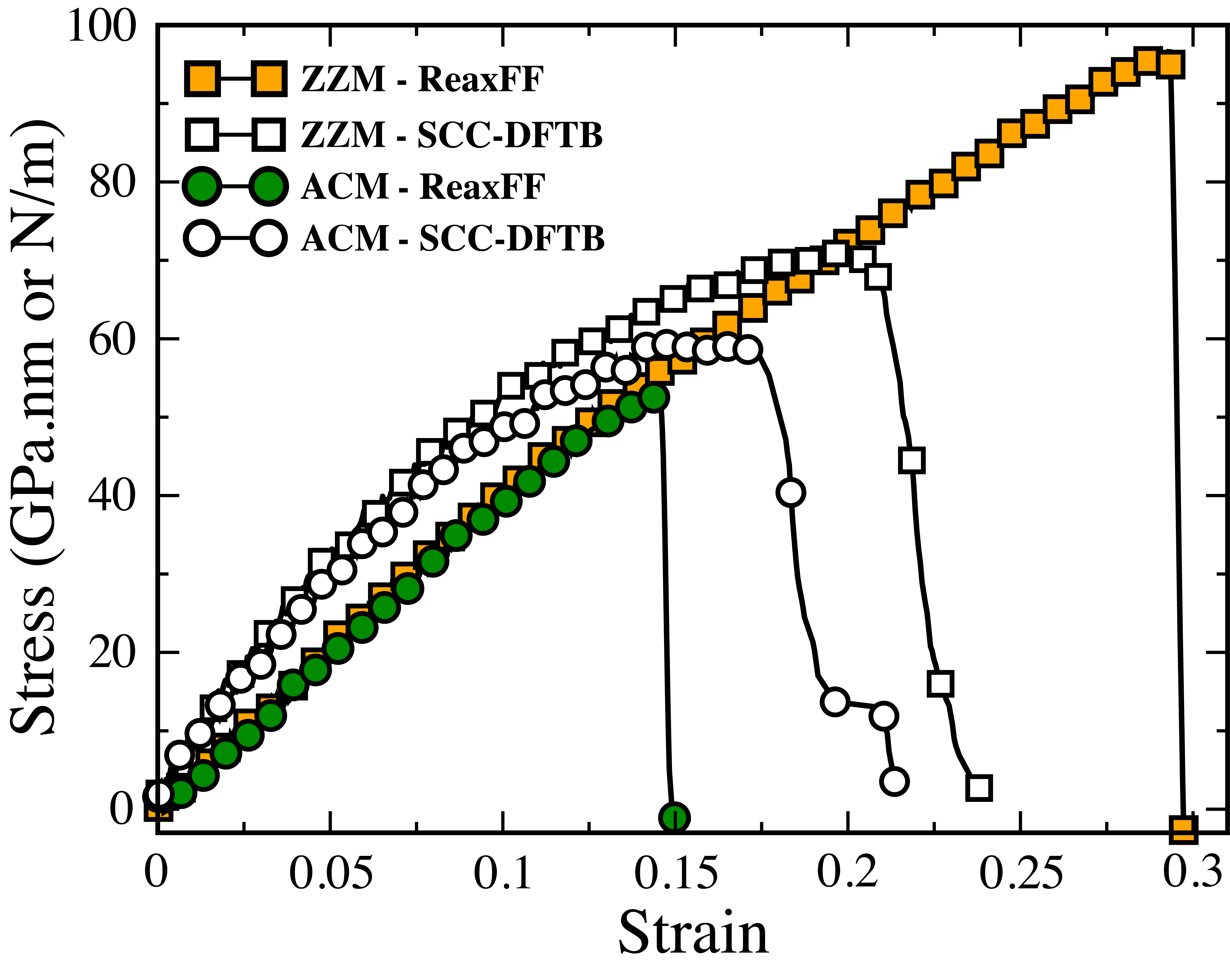}}
\caption{{(Color online) Stress versus strain curves for zigzag and armchair edge terminated structures. Results for the temperature 
of $150$K and for both ReaxFF and SCC-DFTB methods. See text for discussions. 
}}
\label{Fig2}
\end{figure}

We first obtained the minimized geometries for silicene by utilizing the three methods described above: DFT, SCC-DFTB and ReaxFF.  
Graphene and silicene structures, as optimized by the ReaxFF method, can be compared at the same scale as presented in figure \ref{Fig1}. The calculated values for silicene, using the ReaxFF method, were $d=2.3$ \AA ~ for the Si-Si bond length, $\Delta=0.67$ \AA ~ for the buckling value and $\alpha=\beta=112\,^{\circ}$ for the angle value (see figure \ref{Fig1}). DFT and SCC-DFTB calculations resulted, respectively,  in values of $d=2.28$ and  $d=2.32$ \AA ~ for the Si-Si bond length, $\Delta=0.48$ and $\Delta=0.59$ \AA ~ for the buckling and $116\,^{\circ}$ and $113\,^{\circ}$ for both angles $\alpha$ and $\beta$. There is a good agreement between these values and those reported in the literature, see table \ref{table1}. The ReaxFF results for graphene are $d_c=1.42$ \AA ~ for the C-C bond length, no buckling and $\alpha_c=120^{\circ}$ for the bond angle values.

\subsection{Mechanical Properties and Fracture Patterns}

Typical stress-strain curves can be divided into 3 different regions: (i) the harmonic region, where the stress-strain curve is linear and the Young's  Modulus is defined; (ii) the anharmonic region, where the stress increases non-linearly with the increasing strain; and (iii) the plastic region, where the structure undergoes irreversible structural changes. The point at which mechanical failure happens defines two quantities, the final stress, which is the maximum stress value reached before rupturing, and the critical strain $\epsilon_c$, which is the strain value at the moment of rupture. The value of $\epsilon_c$ is taken as the point after which the stress decreases abruptly.

The stress versus strain curves were calculated using both ReaxFF and SCC-DFTB methods, at $150$ K for both zigzag and armchair membranes, as shown in figure \ref{Fig2}. The harmonic region is easily identified as the region where the behavior is linear. This behavior is observed only for sufficiently small strain values and is gradually changed as we move towards the plastic region. As the structure reaches the critical strain value $\epsilon_c$, rupture happens, causing an abrupt fall on the stress values.

Young's Modulus values for armchair and zigzag membranes were obtained by fitting the linear region. We found very small differences between the values for membranes of different edge 
terminations. For the armchair membranes we found the values of $43$ N/m ($0.043$ TPa.nm) with ReaxFF and of $62.7$ N/m ($0.0627$ TPa.nm) with SCC-DFTB. For the zigzag membranes we found the values of $43$ N/m ($0.43$ TPa.nm) with
ReaxFF and $63.4$ N/m ($0.0634$ TPa.nm) with SCC-DFTB. Comparison between the results obtained with SCC-DFTB and values published in the literature shows a very good agreement\cite{PRB2009,PSS2013,RuiQin,topsakalprb,PLA2012}. Young's 
moduli calculated using ReaxFF present a discrepancy of around $30\%$ when compared with these results. However, the qualitative behaviour described by both methods is in very good agrement, as further discussed below. Estimating the thickness of silicene as the van der Waals diameter of $4.2$ \AA\ we obtain a value of $0.149$ TPa for the Young's Modulus in the SCC-DFTB and $0.102$ TPa in the ReaxFF calculations. It is interesting to note that these values are 
7 up to 10 times smaller than the corresponding of graphene ones under similar 
conditions \cite{Ref22,Ref23}. The obtained values for the Poisson ratios were $0.30$ using SCC-DFTB for both ACM and ZZM, and $0.28$ and $0.23$ using ReaxFF for ACM and ZZM, respectively, as shown in Table \ref{table1}.

Despite presenting similar Young's Modulus values, zigzag and armchair membranes exhibit a notable difference in their critical strain values, $\epsilon_c$, as shown in table \ref{table1}. 
The $\epsilon_c$ value is highly dependent on temperature, going from $\epsilon_c=0.20$ and $0.35$ (armchair and zigzag, respectively) at $10$ K to  $\epsilon_c= 0.15$ and $0.30$ at $150$ K. In order to explain this dependence, we stress that kinetic energy fluctuations of atoms in the structure  increase with the temperature. These fluctuations allow the crossing of the energy barrier for the creation of defects at lower strain values. 

There is also a notable dependence on the edge morphology, $\epsilon_c$ differing by a factor of up to 2 if we compare an armchair and a zigzag membrane.
In order to understand this different behaviour of $\epsilon_c$, we have to consider the direction of applied strain in relation to the hexagonal atomic arrangement. 

\begin{figure}[h]
%\vspace*{-1.1cm}
\centerline{\includegraphics[width=1.0\linewidth]{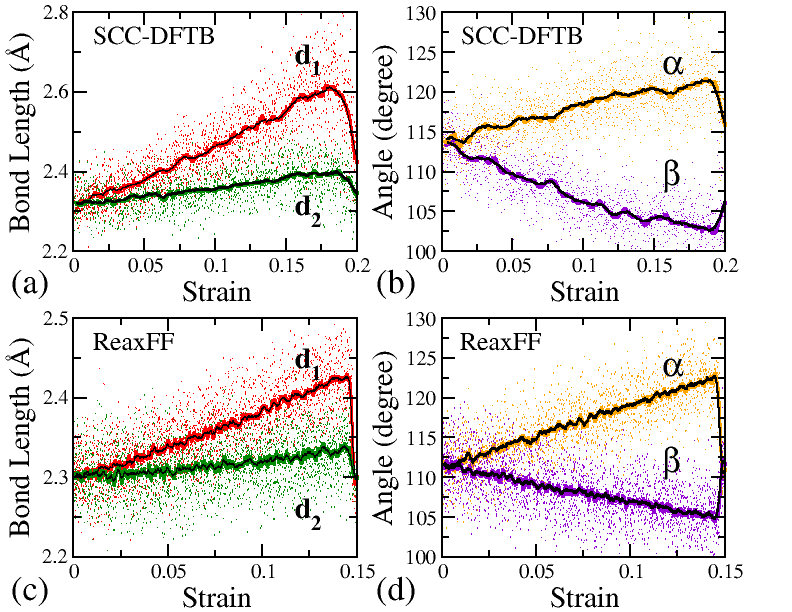}}
\caption{{(Color online) Bonds length and angles values values for ACM. 
$d_1$ is represented by green line, $d_2$ is represented by red line, $\alpha$ orange line and $\beta$ is represented by violet line.}}
\label{Fig4}
\end{figure}
	
\begin{figure}[h]
%\vspace*{-1.1cm}
\centerline{\includegraphics[width=1.0\linewidth]{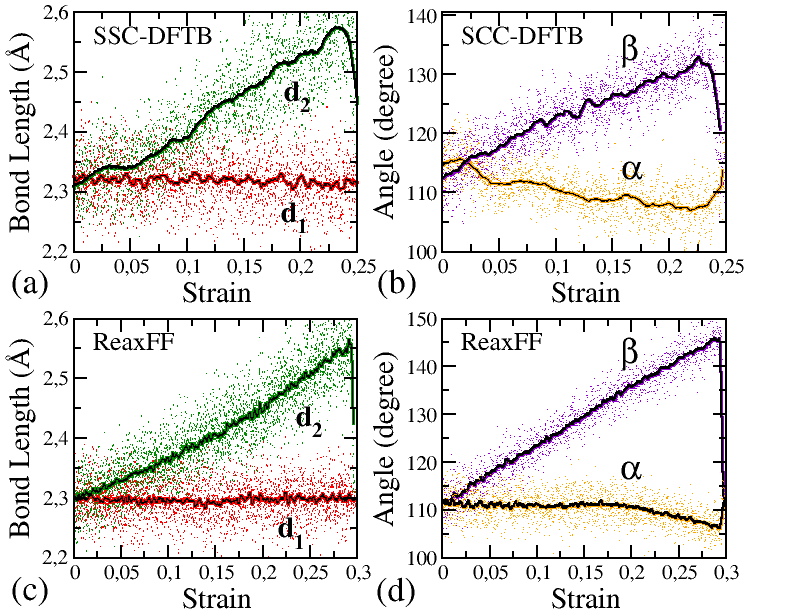}}
\caption{{(Color online) Bonds length and angles values for ZZM. $d_1$ is represented by red line, $d_2$ is represented by green line, $\alpha$ orange line and $\beta$ is represented by violet line. }}
\label{Fig5}
\end{figure}

With the application of strain in the system, the hexagonal symmetry is 
broken and thus two different angles can be defined for each hexagon (figure \ref{Fig1}),  
$\alpha$ and $\beta$, that can either increase or decrease during the deformation process, 
depending on the direction of applied strain. 
As shown in figures \ref{Fig4} and \ref{Fig5}, the dependence of these angles with strain 
is almost linear for $\epsilon < \epsilon_c$. The same symmetry breaking is evidenced by the appearance of two distinct 
bond values, also shown in figures \ref{Fig4} and \ref{Fig5}. 
When strain is applied to armchair membranes, the strain has the same direction of some of the chemical bonds of the structure ($d_1$ as defined in Figure \ref{Fig1}), but this is not true in the case of zigzag membranes. In the latter case, the strain is not parallel to any chemical bond of the structure, so, the relative increase of global strain is not the same as the relative increase of the chemical bond length, while in the case of armchair membranes this can happen for some chemical bonds ($d_1$). This means that, comparing both structures being deformed until they reach the critical chemical bond length value, one can see that zigzag structures must be more strained than their armchair counterparts. This effect redistributes the applied force making zigzag structures more resilient to mechanical deformation.
The curves of the bond lengths versus strain also show clearly the fact that it takes higher strain values for zigzag membranes to reach the same bond lengths as the armchair membranes. This analysis can be extended to graphene as both graphene and silicene share the same honeycomb structure.

\begin{figure}[b]
%\vspace*{-1.1cm}
\centerline{\includegraphics[width=1.0\linewidth]{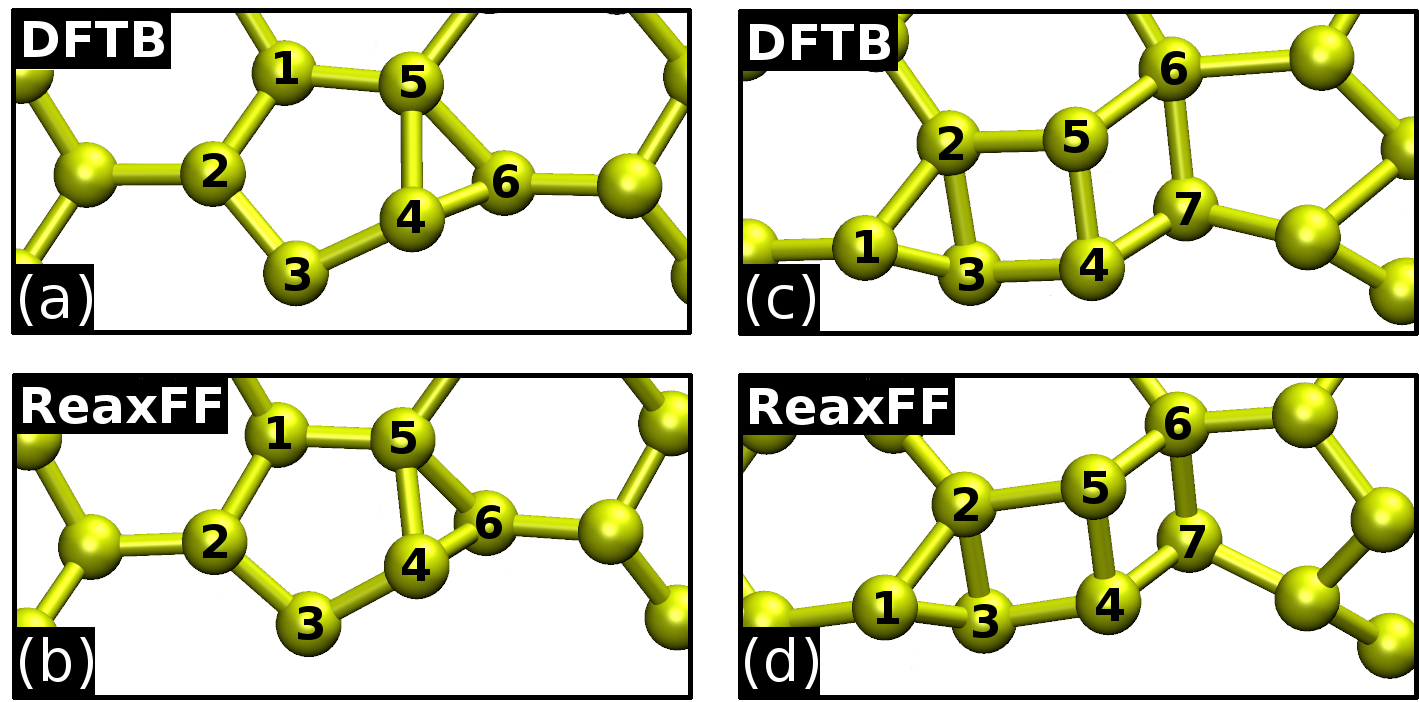}}
\caption{{(Color online)  Detailed view of the edge reconstructions for both ReaxFF and SCC-DFTB methods at (a) and (b) low strain and at (c) and (d) high strain.}}
\label{Fig3}
\end{figure}

\begin{figure}[h!]
%\vspace*{-1.1cm}
\centerline{\includegraphics[width=1.0\linewidth]{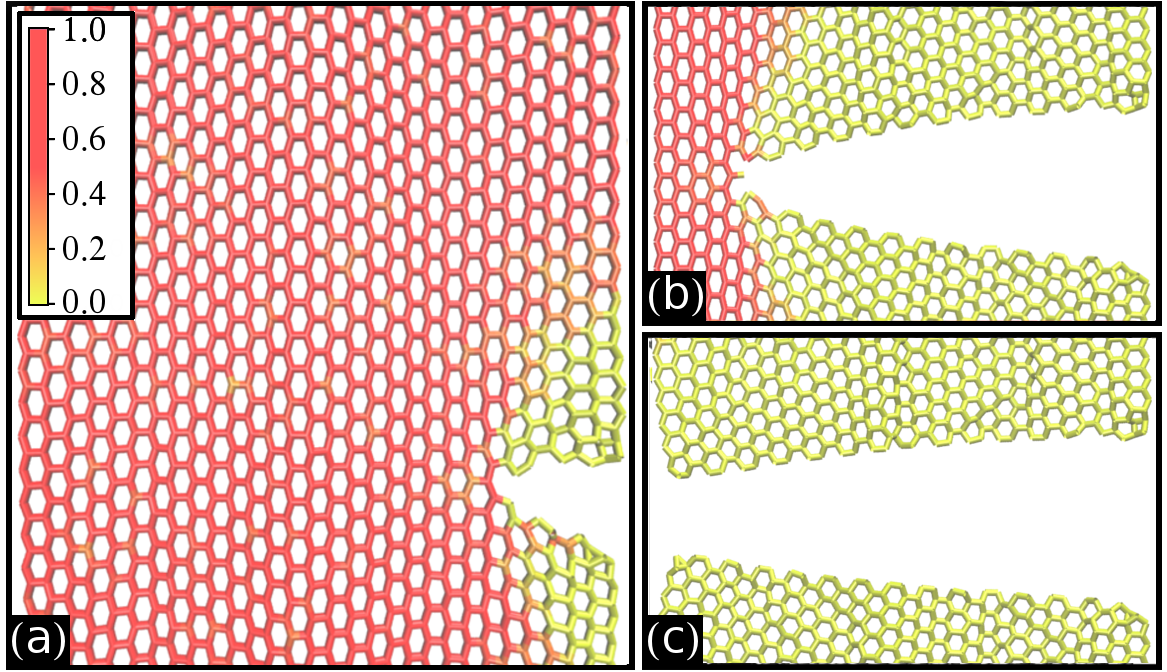}}
\caption{{(Color online)  Typical snapshots from MD simulations showing different stages of the mechanical failure of a zigzag silicene membrane under mechanical strain. The scale goes from low stress (yellow/lighter) to high stress (red/darker).}}
\label{Fig6}
\end{figure}

The stretching dynamics in the plastic region is also dependent on membrane type. 
 For armchair membranes, edge reconstructions are present when it reaches the 
 plastic region. As shown in figure \ref{Fig3} (a) and (b), hexagonal rings are rearranged 
 into pentagonal and triangular ones. Square rings are formed at higher strain levels, as 
 shown in figure \ref{Fig3} (c) and (d). These reconstructions results are consistent for both methods. 
 Triangular and pentagonal rings have been observed in fracture patterns by 
 Topsakal and Ciraci \cite{topsakalprb}. In the case of zigzag membranes no 
 reconstructions were observed (see Figure \ref{Fig6}).

Another unique aspect of silicene under strain is the unbuckling process. We observed the decrease 
of buckling, $\Delta$, with increasing strain, using both methods. This decrease is almost linear 
with angular coefficient of $-0.276$ for armchair and $-0.283$ for zigzag 
using SCC-DFTB and $-1.522$ for both types of membranes using ReaxFF. We observed a continuous 
buckling decrease during the stretching, however, the buckling continues to exist and the structure breaks before its disapperance.

\begin{figure}[h!]
%\vspace*{-1.1cm}
\centerline{\includegraphics[width=1.0\linewidth]{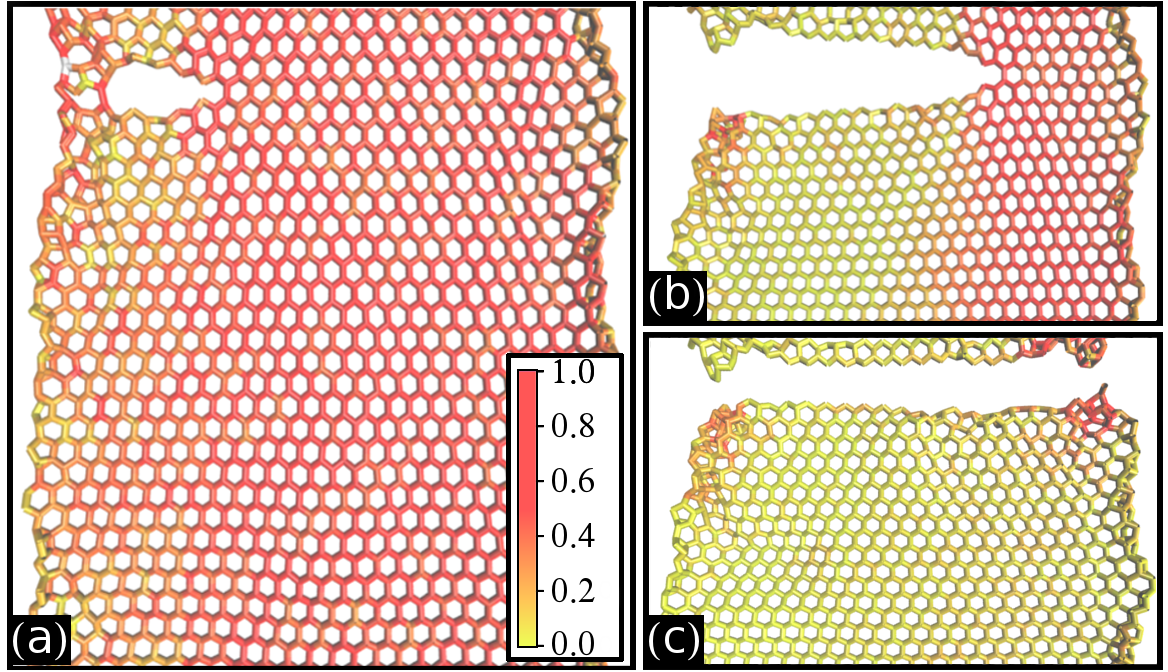}}
\caption{{(Color online) Typical snapshots from MD simulations showing different stages of the mechanical failure of an armchair silicene membrane under mechanical strain. The scale goes from low stress (yellow/lighter) to high stress (red/darker).}}
\label{Fig7}
\end{figure}

We also analysed the von Mises stress distribution, which is defined by equation \ref{vonMises}. Using the ReaxFF method we calculated this distribution along the whole stretching process. Representative snapshots of this process are shown in figures \ref{Fig6} and \ref{Fig7}.
	
For the zigzag membranes the von Misses stress are uniformly distributed before the fracture (figure \ref{Fig6} (a)). When the membrane fracture starts, stress decreases in regions close to the fracture, as shown in figure \ref{Fig6} (b). The rupture creates clean and well-formed armchair edged structures, with only very few pentagon and heptagon reconstructed rings, as shown in figure \ref{Fig6} (c).
	
The corresponding results for the armchair structures present a significant number of edge reconstructions (see Figure \ref{Fig7}(a)), with the formation of mostly pentagon and heptagon rings. As we can see in Figure \ref{Fig7}(b) and (c), in this case the fractured structure presents less clear and more defective zigzag edge terminated structures.
It can also be seen that the von Mises stress distribution is much less uniform during the whole process, even after the fracture starts. This local stress concentration leads to more reconstructed rings in this case. Similar fracture patterns have been observed in graphene \cite{Ref23}, most notably that fractured armchair structures produce zigzag edge terminated ones and vice-versa and with the formation of pentagon and heptagon reconstructed rings.

\section{Summary and Conclusions}
\label{sec5}

We investigated, by means of fully atomistic molecular dynamics simulations under two different methods, ReaxFF and SCC-DFTB, the structural and mechanical properties of single-layer silicene membranes under mechanical strain. There is a qualitative agreement between the results obtained with both methods. Young's modulus values obtained were $43.0$ N/m (for both ACM and ZZM) and $62.7$ N/m (ACM) and $63.4$ N/m (ZZM) using the ReaxFF and the SCC-DFTB methods, respectively. These values present good agreement with those found in the literature. The critical strain and final stress values were shown to be highly dependent on both temperature and edge morphology, the latter being explained by simple geometric arguments.
Temperature also plays a fundamental role in the fracture and reconstruction process. When the system is heated, fracture formation barrier can be transposed and critical strains are lowered. The critical strain value, $\epsilon_c$, goes from $0.20$ and $0.35$ (armchair and zigzag, respectively) at $10$ K to $0.15$ and $0.30$ at $150$ K.

Silicene fracture patterns are similar in some aspects to those observed on graphene, but important differences were also noted, such as, the presence of buckling due to a pseudo 
Jahn-Teller effect. Although the buckling value was progressively reduced during strain application, it was not eliminated, even when significant stress was imposed to the structure, as complete rupture happened before this value could reach zero.

Our results show that, while the Young's moduli values are virtually isotropic for silicene membranes, the critical strain is not. Also, under similar conditions, graphene is many times ($~10$ times) tougher than silicene.

\begin{section}{ACKNOWLEDGMENTS}
This work was supported in part by the Brazilian Agencies, CAPES, CNPq and
  FAPESP. The authors thank the Center for Computational Engineering
  and Sciences at Unicamp for financial support through the
  FAPESP/CEPID Grant \#2013/08293-7. 
  RP acknowledges FAPESP for Grant No. 2013/09536-0. 
  ACTvD acknowledges funding from the U.S. Army Research Laboratory through the Collaborative 
  Research Alliance (CRA) for Multi Scale Multidisciplinary Modeling of Electronic Materials 
  (MSME) and from AFOSR grants  No. FA9550-10-1-0563 and No. FA9550-11-1-0158.

\end{section}

\end{document}